# Multistage ordering and critical singularities in $Co_{1-x}Zn_xAl_2O_4$ ($0 \leq x \leq 1$): Dilution and pressure effects in a magnetically frustrated system

Takashi Naka[1,*] Koichi Sato,[1,2] Yoshitaka Matsushita,[1] Noriki Terada,[1] Satoshi Ishii,[3] Takayuki Nakane,[1] Minori Taguchi,[4] Minako Nakayama,[1] Takeshi Hashishin,[5] Satoshi Ohara,[5] Seiichi Takami,[2] and Akiyuki Matsushita[1]

[1]*National Institute for Materials Science, 1-2-1 Sengen Tsukuba, Ibaraki 305-0047, Japan*
[2]*Institute of Multidisciplinary Research for Advanced Materials, Tohoku University Katahira 2-1-1, Aoba-ku, Sendai 980-8577, Japan*
[3]*Department of Physics, Tokyo Denki University, Hatoyama, Saitama 350-0394, Japan*
[4]*Department of Applied Chemistry, Faculty of Science and Engineering, Chuo University 1-13-27 Kasuga, Bunkyo-ku 112-8551, Japan*
[5]*Joining and Welding Research Institute, Osaka University, 11-1, Mihogaoka Ibaraki, Osaka 567-0047, Japan*

## ABSTRACT

We report comprehensive studies of the crystallographic, magnetic, and thermal properties of a spinel-type magnetically frustrated compound, $CoAl_2O_4$, and a magnetically diluted system, $Co_{1-x}Zn_xAl_2O_4$. These studies revealed the effects of dilution and disorder when the tetrahedral magnetic Co ion was replaced by the nonmagnetic Zn ion. Low-temperature anomalies were observed in magnetic susceptibility at $x < 0.6$. A multicritical point was apparent at $T = 3.4$ K and $x = 0.12$, where the antiferromagnetic, spin glass-like, and paramagnetic phases met. At that point, the quenched ferromagnetic component induced by a magnetic field during cooling was sharply enhanced and was observable below $x = 0.6$. At $x \sim 0.6$, magnetic susceptibility and specific heat were described by temperature power laws, $\chi \sim C/T \sim T^{-\delta}$, in accord with the site percolation threshold of the diamond lattice. This behavior is reminiscent of a quantum critical singularity. We propose an $x$-temperature phase diagram in the range $0 \leq x \leq 1$ for Co$_{1-x}$Zn$_x$Al$_2$O$_4$. The transition temperature of CoAl$_2$O$_4$ determined from magnetic susceptibility measured under hydrostatic pressure increased with increasing pressure.

## I. INTRODUCTION

Systems that exhibit magnetic frustration, such as the spinel-type of cobalt aluminate, $CoAl_2O_4$, have been investigated intensively in the last 10 years. Interest in $CoAl_2O_4$ stems from the fact that the magnetic ground state of the tetrahedrally coordinated $Co^{2+}$ (spin $S = 3/2$) is theoretically highly degenerate as a result of competing magnetic interactions.[1] Because the A-site sublattice forms a diamond lattice known to be magnetically bipartite, the nearest-neighbor antiferromagnetic (AF) interaction, $J_1$, stabilizes an AF state.[2] However, if a next-nearest-neighbor AF interaction, $J_2$, is introduced, frustration effects can be anticipated,[2] because the diamond lattice consists of two interpenetrating, magnetically frustrated fcc lattices shifted along the [111] direction.[2-3] This situation causes strong frustration, which leads to a highly degenerate ground state and a large reduction of the ordering temperature. This ground state degeneracy develops among spin-spiral (SS) states established above $J_2/J_1 = 1/8$, and the corresponding propagation wave vectors form a continuous surface in momentum space.[1] The magnetic ground state changes to an AF state at $J_2/J_1 < 1/8$. Savary *et al.* have considered theoretically the effects of the inversion in frustrated A-site magnetic spinels.[4] $CoAl_2O_4$ is a partially inverted spinel with cobalt ions predominantly in the tetrahedral A-site. A small fraction of cobalt ions occupies the octahedral B-site, and aluminum ions expelled from the B-site consequently occupy tetrahedral A-sites. They have argued that dilute local impurities generally induce an ordered magnetic ground state and act as a degeneracy-breaking force in highly frustrated A-site magnetic spinels.[4] It has also been pointed out that because of the proximity to the Lifshitz point, $J_2/J_1 = 1/8$, the magnetism in $CoAl_2O_4$ is quite sensitive to magnetic and nonmagnetic impurities.[1,4] In contrast, robustness of the SS state against impurities has been demonstrated in $MnSc_2S_4$, which has a larger $J_2/J_1$ of 0.85.[4]

In the ternary spinel oxide system of $DT_2O_4$, where D and T represent divalent and trivalent metallic elements, respectively, the crystal structure (cationic configuration) can be expressed as $(D_{1-\eta}T_\eta)[T_{2-\eta}D_\eta]O_4$ where $(D_{1-\eta}T_\eta)$ and $[T_{2-\eta}D_\eta]$ express the A-site and B-site occupations, respectively, and $\eta$ is the so-called inversion parameter. In polycrystalline $CoAl_2O_4$ specimens examined previously,[2,5,6] cation-site inversions have been unavoidable, despite many efforts to reduce the inversion.[7,8] Recently, a comprehensive investigation[6] has revealed inversion effects on ac- and dc-magnetizations for $(Co_{1-\eta}Al_\eta)[Al_{2-\eta}Co_\eta]O_4$ in the wide range of $0.0467 < \eta < 0.153$. The temperature-$\eta$ phase diagram suggests the possibility of inversion effects on the magnetic state. It should be noted that the assignment of the magnetic ground state is still controversial, as mentioned in Ref. 9. As possible candidates for the ground state of $CoAl_2O_4$, spin glass (SG),[2,6] spin-liquid (SL),[5,6] unconventional (antiferro-)magnetically ordered,[10] and AF[9,11-12] states have been proposed. According to Hanashima *et al.*,[6] at $\eta > 0.08$, the SG state is stabilized below $T_{SG} \sim 4$ K, whereas the SL (or AF) state remains at $\eta < 0.08$. Whereas the anomalous temperature, $T^*$ ($T_N$), decreases rapidly with increasing $\eta$, at the boundary ($\eta \sim 0.08$), the AF (or SL) and SG states coexist (for the definition of $T^*$, see Appendix). As shown below, our specimen of $CoAl_2O_4$ had a relatively small $\eta$ (= 0.055 ± 0.003), at which the AF ground state was realized, as expected. On the basis of the results reported previously and presented in this work, the $\eta$-$T$ phase diagram for a stoichiometric $CoAl_2O_4$ can be constructed (see Fig. A1 in Appendix).

In this paper, we demonstrate the effects of magnetic dilution on the magnetization and specific heat of $Co_{1-x}Zn_xAl_2O_4$ in the wide range of $0 \leq x \leq 1$, and we demonstrate a pressure effect on the transition temperature in $CoAl_2O_4$. Introduction of the nonmagnetic ion $Zn^{2+}$ into the A-site modified the magnetic ground state considerably. The AF state became unstable, and a SG state emerged above $x \sim 0.07$. At $x = 0.12$ and $T = 3.4$ K, a triple critical point was apparent, where the AF, SG, and paramagnetic (PM) phases met. At $x = 0.6$, $\chi(T)$ and $C(T)/T$ seemed to increase monotonically and followed a temperature power law ~ $T^{-\delta}$ below 10 K in Co$_{1-x}$Zn$_x$Al$_2$O$_4$. Additionally, the spin-glass transition seemed to vanish around $x = 0.6$. The implication is that a quantum critical singularity appeared in the vicinity of $x = 0.6$. Actually, a magnetic critical composition was calculated to be $x_P = 0.571$ by employing the nearest-neighbor site percolation model for magnetic A-site spinels. Because the cubic lattice constant of Co$_{1-x}$Zn$_x$Al$_2$O$_4$ decreased with increasing $x$, magnetization of CoAl$_2$O$_4$ was measured under pressure up to 1.3 GPa. The transition temperature of CoAl$_2$O$_4$ increased with increasing pressure.

## II. EXPERIMENTAL

Polycrystalline samples of Co$_{1-x}$Zn$_x$Al$_2$O$_4$ were prepared by solid-solid reaction of the proper amounts of CoO (3N), ZnO (4N), and Al$_2$O$_3$ (4N). The mixed powder was heated at 1300 °C for 24 hours and then cooled to room temperature at a rate of 36 °C/hour. The chemical compositions of the products, evaluated by inductively coupled plasma–atomic emission spectroscopy, were close to the respective nominal compositions. Specifically, the Co/Al and Zn/Al ratios corresponded to the nominal values within an error of 3%. Hereafter, we use the nominal compositions for the crystallographic refinements and analyses of magnetic and thermal properties.

Powder X-ray diffraction measurements using synchrotron radiation ($\lambda = 0.3550$ Å) were conducted on the BL02B2 beam line at SPring-8 (Harima, Japan).[13] The diffraction data revealed no secondary phase, such as CoO, Co$_3$O$_4$, or $\alpha$-Al$_2$O$_3$, in any of the samples. We therefore considered that the samples used in this study consisted of a single phase of Co$_{1-x}$Zn$_x$Al$_2$O$_4$. The crystal structure of the samples was refined with the Rietveld method by using the RIETAN-FP program.[14] On the one hand, we confirmed in the refinement procedures that almost all $Zn^{2+}$ occupied the A-site; the $Zn^{2+}$ ion was therefore fixed on the A-site, without site exchange between $Zn^{2+}$ and $Al^{3+}$. On the other hand, site exchange of $Co^{2+}$ for $Al^{3+}$ was taken into account in the refinements. Therefore, $(Co_{1-\eta-x}Zn_xAl_\eta)[Al_{2-\eta}Co_\eta]O_4$ was adopted as the analytical model in this study. Linear constraints to retain the chemical composition were imposed on the occupancies of the A-site and B-site. The crystallographic parameters obtained from the Rietveld refinements are depicted in Fig. 1. The cubic lattice constant, $a$, decreased at a rate $(1/a)(da/dx) = -2.14 \times 10^{-3}$, whereas the oxygen parameter, $u$, was nearly constant ($u$ = 0.2644–0.2646) with respect to $x$. The inversion parameter, $\eta$, was 0.055(3) for $CoAl_2O_4$ and decreased to 0.0234(23) for $Co_{0.3}Zn_{0.7}Al_2O_4$ with increasing Zn content, $x$. Details of the crystallographic structure and local environment for $Co_{1-x}Zn_xAl_2O_4$ are reported elsewhere.[15]

Dc-magnetization and specific heat down to $T = 2$ K were measured by using a conventional superconducting quantum interference device magnetometer (MPMS-XL; Quantum Design) and a physical property measurement system (Quantum Design), respectively. For high-pressure magnetic measurements, we used a piston-cylinder-type, high-pressure apparatus made from a Cu–Be alloy designed for the magnetometer. Pressure generated at the sample position was calibrated from the pressure-induced variation of the superconducting transition temperature of Sn mounted in a sample holder with the sample.

### III. RESULTS AND DISCUSSION

#### A. $CoAl_2O_4$

##### 1. *Magnetization*

Figure 2(a) shows the temperature dependence of the magnetic susceptibility, $\chi = M/H$, and the reciprocal susceptibility, $1/\chi$, of $CoAl_2O_4$ at a magnetic field strength, $H$, of 10 kOe, where $M$ is the magnetization. Above 100 K, $\chi(T)$ closely followed a modified Curie-Weiss law, that is, $\chi(T) = \chi_0 + C/(T − \theta)$, where $\chi_0$, $C$, and $\theta$ are the temperature-independent susceptibility, the Curie constant, and the Weiss temperature, respectively. For $CoAl_2O_4$, the effective magnetic moment, $\mu_{eff}$ = 4.59 $\mu_B$, and the Weiss temperature, $\theta$ = −91 K, obtained here were in accord with those reported by Tristan et al.[2] and Suzuki et al.,[5] respectively. Here we assumed the constant term in the susceptibility, $\chi_0$, to be the susceptibility of $ZnAl_2O_4$ (−7.4 × 10$^{-5}$ emu/mol).[5,16] At $x$ = 0, a $g$-value of 2.36 was derived from $\mu_{eff}$ based on the assumption that the $S$ = 3/2 spin state for $Co^{2+}$ was located at the A-site.

Note that $\chi(T)$ exhibited a broad maximum at ~14 K [the inset of Fig. 2(b)], consistent with the maxima reported previously.[5,6] The broad observed maximum of $\chi(T)$ is often assumed to be due to short-range order (SRO) above the Néel point in antiferromagnets.[5] Below the maximum of $\chi(T)$, susceptibilities measured after zero-field cooling (ZFC) and field cooling (FC) have been observed to deviate slightly from each other.[2,5] As shown in Fig. 2(b), the derivative with respect to temperature of $\chi(T)$ measured at $H$ = 0.1 kOe, $d\chi/dT$ exhibited a sharp peak. For our sample, the sharp peak in $d\chi/dT$ apparent at $T_{d\chi/dT}$ = 9.0 ± 0.2 K in $CoAl_2O_4$ corresponded to the peak position of specific heat divided by temperature, $C/T$, $T_{C/T}$ = 8.5 ± 0.4 K, as mentioned below (also depicted in Fig. 2(b)). Because the reduction factor, $f = −\theta/T_{d\chi/dT}$ = 10, was comparable to those previously reported,[2,5] a magnetic transition was also strongly suppressed by the bond frustration in our $CoAl_2O_4$ sample.

## *2. Quenched ferromagnetic component*

Interestingly, we found evidence of a quenched ferromagnetic component below 10 K after FC from 30 K to 2 K, as shown in Fig. 3(a). This component depended on the field strength applied during cooling, $H_{FC}$. The samples were field-cooled from $T \gg T_{C/T}$ or $T_{d\chi/dT}$. Figure 3(b) shows the $H_{FC}$ dependence of $\chi$(2K) for $CoAl_2O_4$ measured at $H$ = 0.1 kOe and the field-induced magnetization, $m_0(T, H_{FC})$, defined as $m_0 = [\chi(H_{FC}) − \chi(H_{FC} = 0)]H$. The $\chi(H_{FC})$-temperature curves converged to $\chi(H_{FC} = 0)$ at $T = T_t$, rather higher than $T_{d\chi/dT}$ and $T_{C/T}$. Here, to avoid the effect of the residual magnetic field in the superconducting magnet of the magnetometer, $T_t$ was defined as the temperature where $d\chi(H_{FC})/dT$ converged to $d\chi(H_{FC} = 0)/dT$. [Fig. 3(a)]. We confirmed that $T_t$ was independent of $H_{FC}$ up to $H_{FC} = 50$ kOe. As mentioned above, we could not detect any impurity phases in our samples based on the X-ray diffraction measurements with the synchrotron X-ray source. This small ferromagnetic component would not have been responsible for an undetectable ferromagnetic impurity. Additionally, as can be seen below, $m_0(x)$ showed characteristic behaviors at magnetic phase boundaries apparent in $Co_{1-x}Zn_xAl_2O_4$. It is possible that the field-induced moment arose from cobalt aluminate spinels.

In neutron diffraction measurements on a single crystal of $CoAl_2O_4$,[12] there have been indications of significant field effects on static and dynamic spin correlations. These results have revealed that the static magnetic moment increases, and spin excitations are split by the magnetic field. The dynamic phenomena are presumably due to splitting the manifold of low-lying SL states in $CoAl_2O_4$, which are nearly degenerate in the zero-field state.[12] The significant variation of the magnetic state suggests that applying a magnetic field reveals the proximity of the magnetic state in $CoAl_2O_4$ to the Lifshitz point, $J_2/J_1 = 1/8$. At that point, the spin stiffness, κ, which measures the energetic cost of an infinitesimal change of the spin state, is predicted to vanish.[1] Such a field-induced effect is expected even in polycrystalline samples that are randomly oriented crystallographically with respect to an external magnetic field.

It is worth recalling, as shown in Fig. 3(a), that the termination point, $T_t$, exceeds $T_{d\chi/dT}$ and $T_{C/T}$ and is close to the maximum temperature in $\chi(T)$, where the SRO seemed to develop. The behavior we observed was that the small ferromagnetic moment,ticked was quenched at low temperatures, remained under a small magnetic field of 0.1 kOe even above $T_{d\chi/dT}$, until the temperature reached $T_t$. If we assume that a macroscopic AF ordering occurs at $T_{d\chi/dT}$, it is likely that, when the system is field-cooled, one of the low-lying degenerate states associated with the ferromagnetic component and the termination point $T_t$ can be stabilized in a fragmented condition and embedded into the AF state. In other words, a magnetic phase separation might occur.

## *3. Specific heat*

Anomalous behaviors at relatively low temperatures were also observed in $C/T$, as shown in Fig. 2(b). A peak in $C(T)/T$ was apparent at $T_{C/T} = 8.5 \pm 0.4$ K. Figure 4 shows the magnetic contribution to the specific heat, $\Delta C/T = [C(CoAl_2O_4) - C(ZnAl_2O_4)]/T$. The magnetic contribution $\Delta C/T$ followed the temperature power law $\Delta C \sim T^{1+\alpha}$ below $T_{C/T}$. The exponent 1 + $\alpha$ equaled 2.33. This exponent was comparable with those previously reported for $CoAl_2O_4$ samples.[2,5,9] The normalized magnetic entropy, $S_{mag}(T)/(R \ln 4)$, calculated from $\Delta C(T)/T$ was in good agreement with previously reported values.[2,5,9] The magnetic entropies at $T = T_{C/T}$ and $T_t$ were quite small: $S_{mag}(T)/(R \ln 4) = 0.27$ and 0.45, respectively. As previously pointed out,[9] 75% of the full magnetic entropy is released above $T = T_{C/T}$. This fact is consistent with the evidence of SRO above $T_{d\chi/dT}$ (or $T_{C/T}$) in $\chi(T)$, as mentioned above. These results, obtained for our $CoAl_2O_4$ with $\eta = 0.055(3)$, are in accord with those in the recent investigation by Roy et al.[9] They also conducted neutron diffraction studies with $^{59}Co$ NMR and $^{27}Al$ NMR on a polycrystalline $CoAl_2O_4$ sample with a similar value of the inversion, $\eta = 0.057(20)$, and they confirmed that an AF ordering occurs below $T_N = 9.8$ K. Significantly, a dynamic SRO was observed well above $T_N$. By analyzing the momentum transfer ($q$) dependence of the elastic neutron scattering intensity, they concluded that the SRO results mainly from an inversion effect.

## B. $Co_{1-x}Zn_xAl_2O_4$

### *1. Magnetization*

Figure 5(a) shows the temperature dependence of the magnetic susceptibility of $Co_{1-x}Zn_xAl_2O_4$ at $H = 10$ kOe. Above 100 K, $\chi(T)$ followed the Curie-Weiss law [the inset of Fig. 5(a)]. As in the case of $CoAl_2O_4$, we assumed the constant term in the susceptibility, $\chi_0$, to be the susceptibility of $ZnAl_2O_4$ and to be independent of $x$ in the range of $0 < x \leq 1$. As shown in Fig. 5(b), the $g$-value derived from $\mu_{eff}(x)$ decreased with increasing $x$. As mentioned previously, a decrement of the local Co–O distance, <Co–O>, with increasing $x$ was apparent and likely caused the increment of the crystal field splitting, $\Delta$, because it is expected that $\Delta \propto 1/<Co–O>^5$ (Ref.17). In addition, because of a reduction of the effective spin-orbit coupling, $\lambda'$, due to the enhancement of the orbital quenching of the Co ion by the degree of covalence of the Co–O bond, the $g$-value, $g \approx g_0 + \lambda'/\Delta$, of the Co ion decreases, where $g_0 = 2.0$ is the $g$-value for the free spin.[18] A linear variation of $\theta$ with respect to $x$ was apparent at $0 < x < 1$ [Fig. 5(b)].

As can be seen in Fig. 6(a), the low-temperature behavior of $\chi(T)$ varied strongly with increasing $x$. In view of the AF long-range order established for $CoAl_2O_4$ with $\eta = 0.05$–0.06, it is reasonable to relate the $x$-variation of the AF transition with the peaks of $d\chi(T)/dT$ and $C(T)/T$ for $Co_{1-x}Zn_xAl_2O_4$. The peak position of $d\chi(T)/dT$, $T_{d\chi/dT}$, obtained at $H = 0.1$ kOe decreased rapidly with increasing $x$ [Fig. 6(b)]. That is, the magnetic ordering was suppressed strongly with an increase of the Zn content at the A-site. At $x = 0.075$–0.30, the low-field susceptibility, $\chi(T)$, after ZFC showed a cusp at $T = T_{cusp}$ [Figs. 6(a) and 7(a)–(b)]. Below $T \sim T_{cusp}$, the discrepancy between the $\chi$-$T$ curves after ZFC and FC became significant at $0.1 < x < 0.4$, but this feature smeared out above $x \sim 0.4$. The cusp in $\chi(T)$ observed at $x \geq 0.075$ could signal a SG transition, as indicated previously in $(Co_{1-\eta}Al_\eta)[Al_{2-\eta}Co_\eta]O_4$.[2,6] It should be emphasized that at $x = 0.075$–0.10, a multiple transition occurs at $T = T_{cusp}$ and $T_{d\chi/dT}$, as indicated in Figs. 6(a)–(b). This transition was also observed in $(Co_{1-\eta}Al_\eta)[Al_{2-\eta}Co_\eta]O_4$ at $\eta = 0.079$–0.085. The termination point $T_t(x)$ decreased with increasing $x$ and vanished as expectedly at $x \sim 0.6$, but it remained higher than $T_{C/T}$ and $T_{cusp}$ below $x \sim 0.6$, as shown in Figs. 7(a) and 7(c).

## 2. Specific heat

The peak in the $C/T$ curve became indistinct, and the peak temperature, $T_{C/T}$, decreased with increasing $x$. Figure 8(a) shows the magnetic contribution to specific heat, $\Delta C/T = [C(x) - C(ZnAl_2O_4)]/T$. In $Co_{1-x}Zn_xAl_2O_4$, a broad maximum or plateau region in $C/T$, apparent above $T_{cusp}$, was observable above $x \sim 0.1$. The rapid broadening of $C(T)/T$ with increasing $x$ was reflected in the multiple transition apparent in $\chi(T)$ at $x = 0.075$–0.10, and the SG transition stabilized above $x \sim 0.1$. Generally, a broad maximum in specific heat is observed slightly above the SG transition.[19] In this study, we observed that with increasing $x$, $T_{C/T}(x)$ deviated from $T_{d\chi/dT}(x)$, as shown in Figs. 6(b) and 8(a). Therefore, determining a magnetic transition temperature at $x > 0$ from the $\Delta C(x, T)/T$ curves was difficult. At $x = 0$–0.10, the magnetic contribution $\Delta C/T$ followed the temperature power law $\Delta C \sim T^{1+\alpha}$ below $T_{C/T}$. The exponent $1 + \alpha$ equaled 2.1 for $x = 0.05$ and 1.9 for 0.10.

In contrast with the multiple transition and the SG regions, at $x \sim 0.6$, $\Delta C(T)/T$ increased rapidly with decreasing temperature and obeyed a power law, $\sim T^{-\delta}$, below $T \sim 10$ K, which seems to be a ubiquitous singularity observed in the vicinity of the magnetic quantum critical point, $T_{C,N} \to 0$.[20-22] Accordingly, in the relatively low-temperature region, $T < |\theta|$, $\chi(T)$ also obeyed a power law at $x = 0.6$ [Fig. 8(b)]. In this magnetically diluted system, these singularities are consistent with the fact that the nearest-neighbor site percolation threshold of the diamond lattice is estimated to be $y_p = 0.429$.[23] Note that $x_p = 1 - y_p = 0.571$ in the notation of $Co_{1-x}Zn_xAl_2O_4$. Coincidently, as mentioned above, the field-quenched moment, $m_0$, was detected at $x < 0.6$, but not at $x \geq 0.6$ [Fig. 9(b)]. Although the experimental range should be expanded to lower temperatures to confirm the singular temperature dependence, $\sim T^{-\delta}$, with decreasing temperature at $x \sim x_c$, the obtained exponent of $\Delta C/T$ corresponded well with that of $\chi$ [Fig. 8(b)]. As can be seen in Fig. 10(a), in $Co_{1-x}Zn_xAl_2O_4$ a broad maximum or plateau region apparent above $T_{cusp}$ in $C/T$ seemed to converge on $T = 0$ K at $x \sim 0.6$. Likewise, the anomalous points in $(T/\chi)(d\chi/dT)$ corresponding to $T_{cusp}(x)$ and $T_t(x)$ diminished at $x = 0.5$–$0.7$ [Figs. 10(b)–(c)]. Consequently, we expected that the SG phase would disappear at $x \sim 0.6$, and a PM state would emerge above $x \sim 0.6$.

### 3. x-T phase diagram and field-quenched component

The characteristic temperatures, $T_{d\chi/dT}$, $T_{cusp}$, and $T_t$ obtained at $H = 0.1$ kOe are depicted as a function of $x$ in Fig. 9(a). The AF-PM critical phase boundary, $T_{d\chi/dT}(x)$, and the SG-PM boundary, $T_{cusp}(x)$, met at the multicritical point of $T_{tc} = 3.4$ K, $x_{tc} = 0.12$. Correspondingly, as can be seen in Fig. 9(b), a sharp peak in $m_0(2K)$ was observed at $x = 0.12$, although thermodynamic quantities such as $C/T$ and $\chi$ showed no evidence of singular behavior. $T_{cusp}$ seemed to appear at $x = 0.05$–$0.07$, reached a maximum at $x = 0.15$, and decreased gradually with increasing $x$.

Multicritical behavior and its first-order characteristics were apparent from a theoretical standpoint as follows. In the theoretical $(T/J_1)$-$(J_2/J_1)$ phase diagram,[1] the AF-PM transition temperature decreased sharply with increasing $J_2$. Just above $J_2/J_1 = 1/8$, two magnetic transitions appeared, the PM-AF and the AF-SS transitions. The AF state slightly above $J_2/J_1$

= 1/8 was thermally stabilized ("thermal order-by-disorder" [4]); whereas at $T = 0$ the AF state diminished at $J_2/J_1 = 1/8$, and the SS state arose above $J_2/J_1 = 1/8$. Throughout this region, the transition was expected to be first order. It should be noted that massive degeneracy occurred at a relatively low temperature, even around the triple critical point in $Co_{1-x}Zn_xAl_2O_4$, because at $x \sim 0.1$ the ordering temperature was greatly lowered by the Zn substitution compared with $\theta(x)$ [Fig. 5(b)].

The $x$-$T$ phase diagram in Fig. 9(a) is similar to the $\eta$-$T$ phase diagram presented in Ref. 6 (see also Fig. A1 in Appendix). There were transition regions from the AF to SG states located at $x \sim 0.10$ and $\eta \sim 0.08$ in the $x$-$T$ and $\eta$-$T$ diagrams, respectively. Also, multiple magnetic transitions were observed in the transition region in both cases. For the former, both of the nonmagnetic ions, $Zn^{2+}$ and $Al^{3+}$, occupied the A-site. At the transition of $x = 0.1$, slightly lower than $x_{tc} = 0.12$, the occupation number of the nonmagnetic ions in the A-site of $Co_{1-x}Zn_xAl_2O_4$ was $\eta + x \sim 0.15$ larger than that ($\eta = 0.08$) observed in $CoAl_2O_4$.[6] This difference means that the effect of the Zn substitution on the AF transition was moderate compared with the inversion effect. This difference was probably due to a strong magnetic coupling between the magnetic ions, $Co^{2+}$(A-site) and $Co^{2+}$(B-site),[4,6] and the decrement of $\eta$ with increasing $x$ in $Co_{1-x}Zn_xAl_2O_4$.

At $x \sim 0.6$, a quantum magnetic transition seemed to take place from the SG to the PM state. At $x \sim 0.1$, spin fluctuations were not developed strongly, whereas at $x \sim 0.6$, thermodynamic quantities were strongly enhanced by the magnetic fluctuations as $T \rightarrow 0$ according to Fig. 8(b). Note that at these points, $m_0(x)$ showed characteristic behavior: it increased sharply at $x = 0.12$ and disappeared above $x \sim 0.6$ [Fig. 9(b)].

### *4. Pressure effect on the magnetic transition in CoAl₂O₄*

The AF state of $CoAl_2O_4$ was stabilized under high pressure, while the AF state of $Co_{1-x}Zn_xAl_2O_4$ became unstable with increasing $x$. Note that $a$ decreases with increasing $x$ in $Co_{1-x}Zn_xAl_2O_4$, as shown above. Figure 11(a) shows $d\chi(T)dT$ at various pressures in $CoAl_2O_4$. $T_{d\chi/dT}$ increased with increasing pressure at a rate $(1/T_{d\chi/dT})(dT_{d\chi/dT}/dP) = 0.044$ GPa$^{-1}$ [Fig. 11(b)]. This rate is comparable to $(1/T_N)(dT_N/dP) = 0.071$ GPa$^{-1}$ for the A-site AF spinel oxide $Co_3O_4$ with $T_N \sim 30$ K and $a = 8.084$ Å at ambient pressure.[24] The cubic lattice constant at the highest pressure (= 1.3 GPa), estimated to be 8.085 Å from the bulk modulus $B_0 = 200$ GPa,[8] was comparable to that of $Co_3O_4$ at ambient pressure. However, the observed transition temperature, $T_{d\chi/dT}$ (1.3 GPa) = 9.4 K, was much smaller than $T_N$ ($P = 0$) ~ 30 K for $Co_3O_4$. The positive correlation with pressure was opposite that observed for $T_{d\chi/dT}(x)$ in the case of the Zn-substituted system. To compare both effects quantitatively on the transition temperature, $T_m$, we can introduce a volume coefficient, $r_v = d(\ln T_m)/d(\ln V)$, and the relations, $r_v = -B_0(1/T_m)(dT_m/dP) = (1/T_m)(dT_m/dx)/3r_x$, where $r_x$ is $(1/a)(da/dx)$, estimated above to be $-2.14 \times 10^{-3}$ for $Co_{1-x}Zn_xAl_2O_4$. The volume coefficients were estimated to be $-8.8$ and $950$ from the pressure and the Zn substitution effects on $T_{d\chi/dT}$, respectively. The apparent large discrepancy and also opposite sign of the volume coefficients are important facts to consider with respect to the $x$-variation of the magnetic exchange interactions in $Co_{1-x}Zn_xAl_2O_4$ and the proximity to the Lifshitz point in $CoAl_2O_4$, where there is expected to be the magnetic phase boundary between the AF and the SS states.

Although the ionic radius of $Zn^{2+}$ in tetrahedral coordination (0.60 Å) is larger than that of $Co^{2+}$ (0.58 Å),[25] the lattice constant decreases with increasing $x$. From the lattice constant of $ZnAl_2O_4$ [= 8.08598(1) Å] and the oxygen parameter [= 0.26455(8)] obtained from the structure refinement,[15] we estimated the Zn–O bond length to be 1.9544(4) Å. On the basis of this value and the ionic radius of $O^{2-}$ (= 1.38 Å[25]), the effective ionic radius of $Zn^{2+}$ in $ZnAl_2O_4$ was estimated to be 0.57 Å. A further increase of the extent of Zn substitution shortened the Co–O bond in $Co_{1-x}Zn_xAl_2O_4$.[17] In fact, there were considerable effects due to the nature of the A–O bond on the optical absorption intensity of the d-d transition for $Co^{2+}$ ion[17,26] and the $g$-value [Fig. 2(b)] with increasing $x$ in $Co_{1-x}Zn_xAl_2O_4$. It is possible that variations of the covalency and the metal-oxygen bond lengths affected the average value of $J_2/J_1$ and, consequently, the magnetic ground state. We can therefore ask whether the significant multistage ordering, predicted theoretically, is relevant to the ordering and the tricritical point observed experimentally in $Co_{1-x}Zn_xAl_2O_4$.

## IV. CONCLUSION

We demonstrated the effects of magnetic dilution on crystal structure, magnetization, and specific heat, and we proposed an $x$-$T$ magnetic phase diagram for $Co_{1-x}Zn_xAl_2O_4$ in the range $0 \leq x \leq 1$. The field-induced ferromagnetic component observable below $x = 0.6$ was enhanced at $x = 0.12$ in the vicinity of the multicritical point $(x_{tc}, T_{tc}) = (0.12, 3.4 \text{ K})$, where the AF, SG, and PM phases met. At the nearest-neighbor site percolation threshold, $x_p$, of the diamond lattice consisting of the A-site in the spinel structure, the susceptibility, $\chi$, and $C/T$ seemed to increase rapidly with decreasing temperature and followed a temperature power law $\sim T^{-\delta}$, the suggestion being that these quantities exhibit a quantum singularity associated with strong magnetic fluctuations. It is generally expected that the magnetic state and the value of $J_2/J_1$ can be tuned continuously by chemical substitutions or by application of external mechanical pressure. The transition temperature $T_{d\chi dT}$ of $CoAl_2O_4$ determined from the magnetic susceptibility measured under hydrostatic pressure increased with increasing pressure at a rate $(1/T_{d\chi dT})(dT_{d\chi dT}/dP) = 0.044 \text{ GPa}^{-1}$.

## APPENDIX

Figure A1 shows the anomalous points observed in magnetic susceptibility, $\chi(T)$, and specific heat, $C(T)$,[2,5-6,9-12] and nuclear magnetic resonance (NMR)[9] plotted as a function of $\eta$. The PM state is separated from the AF and the SG states by phase boundaries that consist of the SG-like transition temperature, $T_{SG}(\eta)$, and $T_N(\eta)$. $T_{SG}$ is defined as the temperature at which $\chi(T)$ exhibits a cusp measured after ZFC. $T_N$ can be determined from $T_{C(T)}$, $T_\chi$, and $T_{d\chi/dT}$, which are defined to be the temperatures associated with the maxima of $C(T)$, $\chi(T)$, and the derivative with respect to temperature of $\chi(T)$, $d\chi/dT$, respectively. Roy *et al.* determined precisely the Néel point, $T_N$(NMR), from the line width of the $^{27}$Al NMR, $W(T)$, within ±0.2 K for a polycrystalline sample with $\eta = 0.057$.[9] Hanashima *et al.* have defined $T_\chi$ (or $T^*$ denoted in Ref. 6) to be the temperature at which susceptibilities measured after ZFC and FC deviate from each other.[6] The anomalous points reported in the previous studies and in this work seem to collapse into the phase boundary obtained by Hanashima *et al.*,[6] as shown in Fig. A1. In the case of the smallest $\eta$ (= 0.02 ± 0.04) reported until now of a single-crystal sample used for a neutron diffraction measurement,[11] the anomalous points are 6 and 7.2 K for $\chi$ and $C/T$, respectively. It would be quite interesting to investigate whether $T_N$ is suppressed as $\eta$ approaches 0.

## ACKNOWLEDGEMENTS

We thank Mr. K. Kitahara, Mr. S. Hishikawa, and Professor C. Numako (Chiba University, Japan) for assistance with the synchrotron X-ray diffraction experiments conducted at SPring-8. We appreciate the help of the Material Analysis Station of the National Institute for Materials Science for the inductively coupled plasma–mass spectrometry analyses. This work was partially supported by a Grant-in-Aid for Scientific Research, KAKENHI, (25246013) and by the Network Joint Research Center for Materials and Devices. The synchrotron radiation experiments were performed on the BL02B2 beam line at SPring-8, with the approval of the Japan Synchrotron Radiation Research Institute (2013A1569).

**References**

[1] D. Bergman, J. Alicea, E. Gull, S. Trebst and L. Balents, Nature Physics, **3**, 487-491 (2007).

[2] N. Tristan, J. Hemberger, A. Krimmel, H-A. Krug von Nidda, V. Tsurkan, and A. Loidl, Phys. Rev. B, **72**, 174404 (2005).

[3] V. Fritsch, J. Hemberger, N. Buttgen, E.-W. Scheidt, H.-A. Krug von Nidda, A. Loidl, and V. Tsurkan: Phys. Rev. Lett. **92**, 116401 (2004).

[4] L. Savary, E. Gull, S. Trebst, J. Alicea, D. Bergman and L. Balents, Phys. Rev. B **84**, 064438 (2011).

[5] T. Suzuki, H. Nagai, M. Nohara and H. Takagi, J. Phys.: Condens. Matter, **19**, 145265 (2007).

[6] K. Hanashima, Y. Kodama, D. Akahoshi, C. Kanadani, and T. Saito, J. Phys. Soc. Jpn., **82**, 024702 (2013).

[7] A. Nakatsuka, Y. Ikeda, Y. Yamasaki, N. Nakayama and T. Mizota, Solid State Commun. **128**, 85 (2003).

[8] F. Tielens, M. Calatayud, R. Franco, J. M. Recio, J. Pérez-Ramírez, and C. Minot, J. Phys. Chem. B, **110**, 988 (2006).

[9] B. Roy, Abhishek Pandey, Q. Zhang, T. W. Heitmann, D. Vaknin, D. C. Johnston and Y. Furukawa, Phys. Rev. B, **88**, 174415 (2013). They listed magnetic transition temperature, Weiss temperature, frustration parameter, effective magnetic moment ($g$-value) and assignment of the magnetic ground state reported in literatures[2,5-6,10-12] and in their work[9] as a function of η.

[10] G. J. MacDougall, D. Gouta, J. L. Zarestky, G. Ehlers, A. Podlesnyak, M. A. McGuire, D. Mandrus, and S. E. Nagler, PNAS, **108**, 15693–15698 (2011). See also the supplemental information of this article.

[11] A. Maljuk, V. Tsurkan, V. Zestrea, O. Zaharko, A. Cerellino, A. Loidl, and D. N. Argyriou, J. Cryst. Growth. 311, 3997 (2009).

[12] O. Zaharko, N. B. Christensen, A. Cervellino, V. Tsurkan, A. Maljuk, U. Stuhr, C. Niedermayer, F. Yokaichiya, D. N. Argyriou, M. Boehm, and A. Loidl, Phys. Rev. B, **84**, 094403 (2011).

[13]E. Nishibori, M. Takata, K. Kato, M. Sakata, Y. Kubota, S. Aoyagi, Y. Kuroiwa, M. Yamakata, and N. Ikeda, Nuc. Instrum. Methods Phys. Res. A, **467-468**, 1045 (2001).

[14]F. Izumi and K. Momma, Solid. State Phenom., **130**, 15 (2007).

[15]K. Sato et al. in preparation.

[16]P. Cossee and A. E. van Arkel, J. Phys. Chem. Solids, **15**, 1 (1960).

[17]M. Ardit, G. Cruciani and M. Dondi, Am. Mineral. **97**, 1394 (2012).

[18]J. Owen and J. H. M. Thornley, Rep. Prog. Phys. **29**, 675 (1966).

[19]See for example: L. E. Wenger, P. H. Keesom, Phys. Rev. B, **11**, 3497 (1975).

[20]D. S. Fisher, Phys. Rev. Lett. **96**, 534 (1992).

[21]C. Pich, A. P. Young, H. Rieger and N. Kawashima, Phys. Rev. Lett., **81**, 5916 (1998).

[22]O. Motrunich, S.-C. Mau, D. A. Huse and D. S. Fisher, Phys. Rev. B, **61**, 1160 (2000).

[23]F. Scholl and K. Binder, Zeitschrift für Physik **B39** 239 (1980).

[24]Y. Ikedo, J. Sugiyama, H. Nozaki, K. Mukai, H. Itahara, P. L. Russo, D. Andreica, A. Amato, Physica B, **404**, 652 (2009).

[25]R. D. Shannon, Acta. Cryst. Sec. A, **32**, 751 (1976).

[26]T. Nakane, T. Naka, K. Sato, M. Taguchi, M. Nakayama, T. Mitsui, A. Matsushita and T. Chikyow, Dalton Trans., **44**, 997 (2015).

**Figures and figure captions**

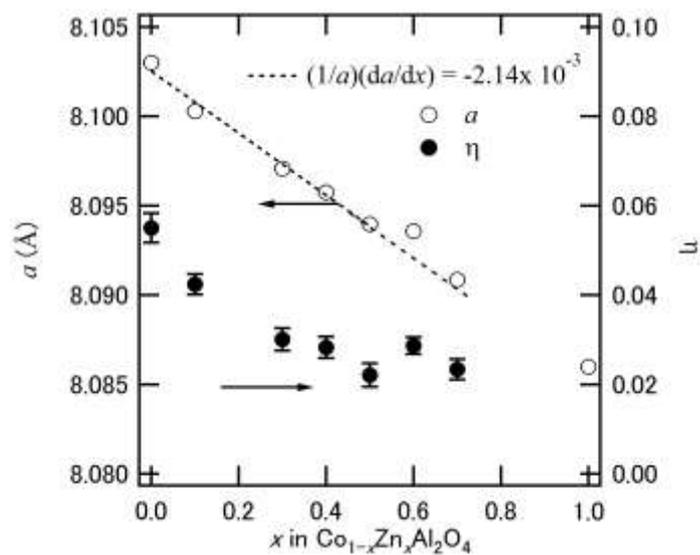

FIG. 1. The *x*-variations of the lattice constant (open circle) and inversion parameter (filled circle) for $Co_{1-x}Zn_xAl_2O_4$ at room temperature. The error bar of the lattice constant is within the symbol. For $0 < x < 1$, it is assumed that the Zn ion dominates only at the A-site.

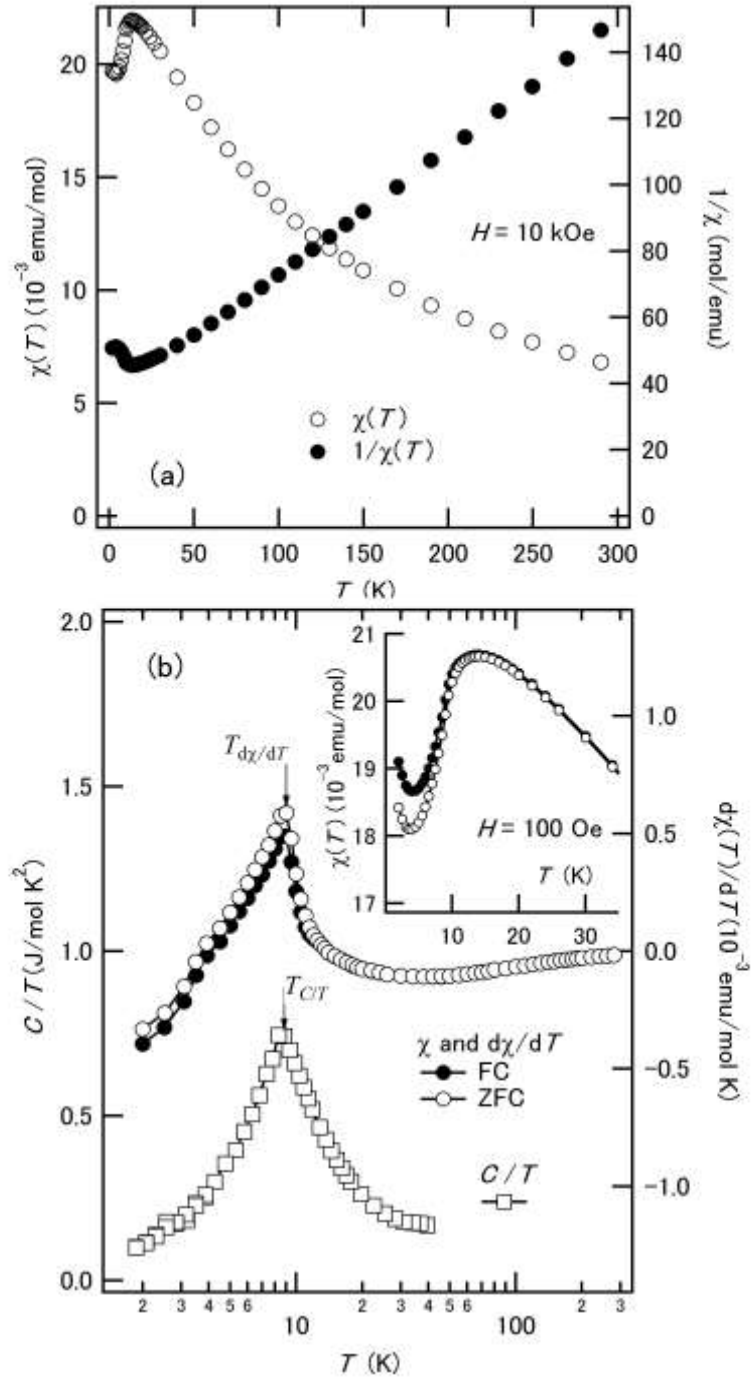

FIG. 2. (a) Magnetic susceptibility and reciprocal susceptibility of $CoAl_2O_4$ as a function of temperature measured at $H = 10$ kOe. (b) $d\chi/dT$ obtained at the applied magnetic field of $H = 0.1$ kOe and $C/T$ as a function of temperature. The inset shows the temperature dependence of the magnetic susceptibility at lower temperature. Arrows indicate the peak temperatures in $d\chi(T)/dT$ and $C(T)/T$.

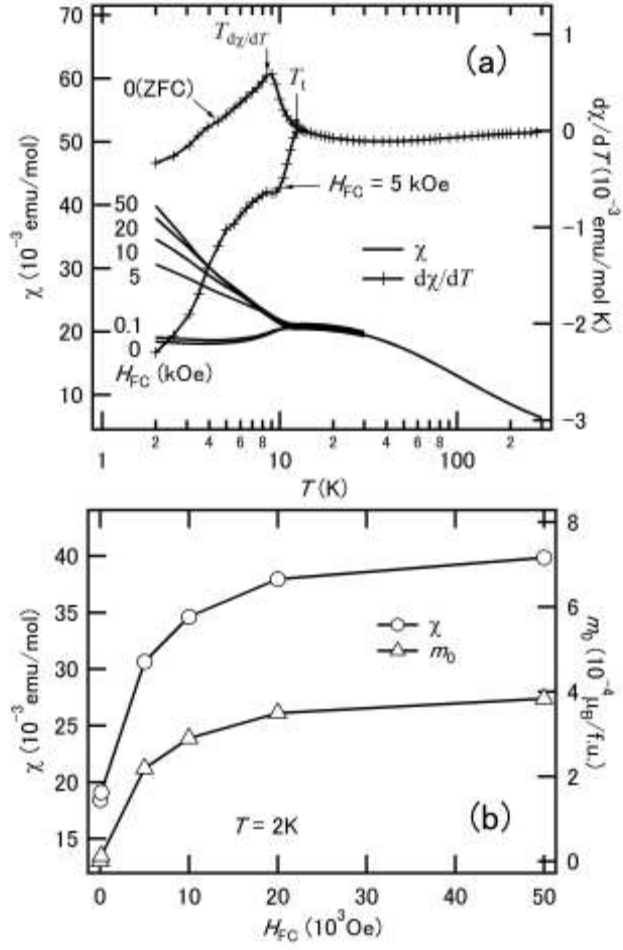

FIG. 3. (a) The $H_{FC}$ dependences of $d\chi/dT$ and $\chi$ for $x = 0$ measured in $H = 0.1$ kOe. Vertical arrows indicate $T_{d\chi/dT}$ and the termination point of $T_t$, respectively. (b) Magnetic susceptibility and the induced moment, $m_0$, for $x = 0$ measured at $T = 2$ K as a function of $H_{FC}$.

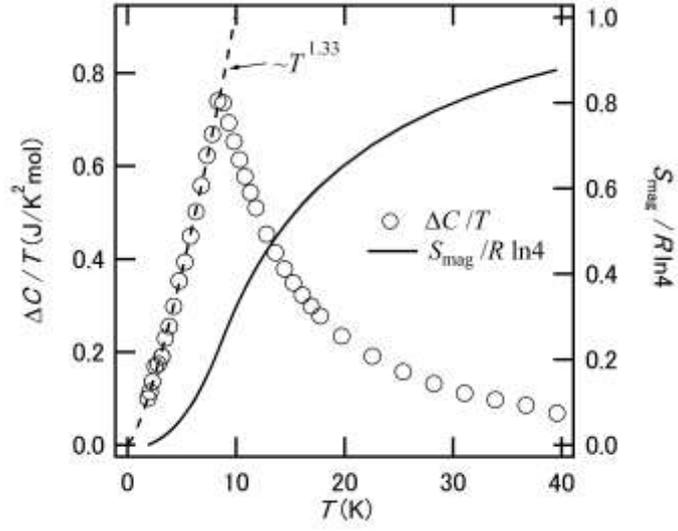

FIG. 4. Magnetic contribution to specific heat and the normalized magnetic entropy for $CoAl_2O_4$ as a function of temperature measured at $H = 0$. The dashed line represents a power-law temperature dependence established below $T_{C/T}$.

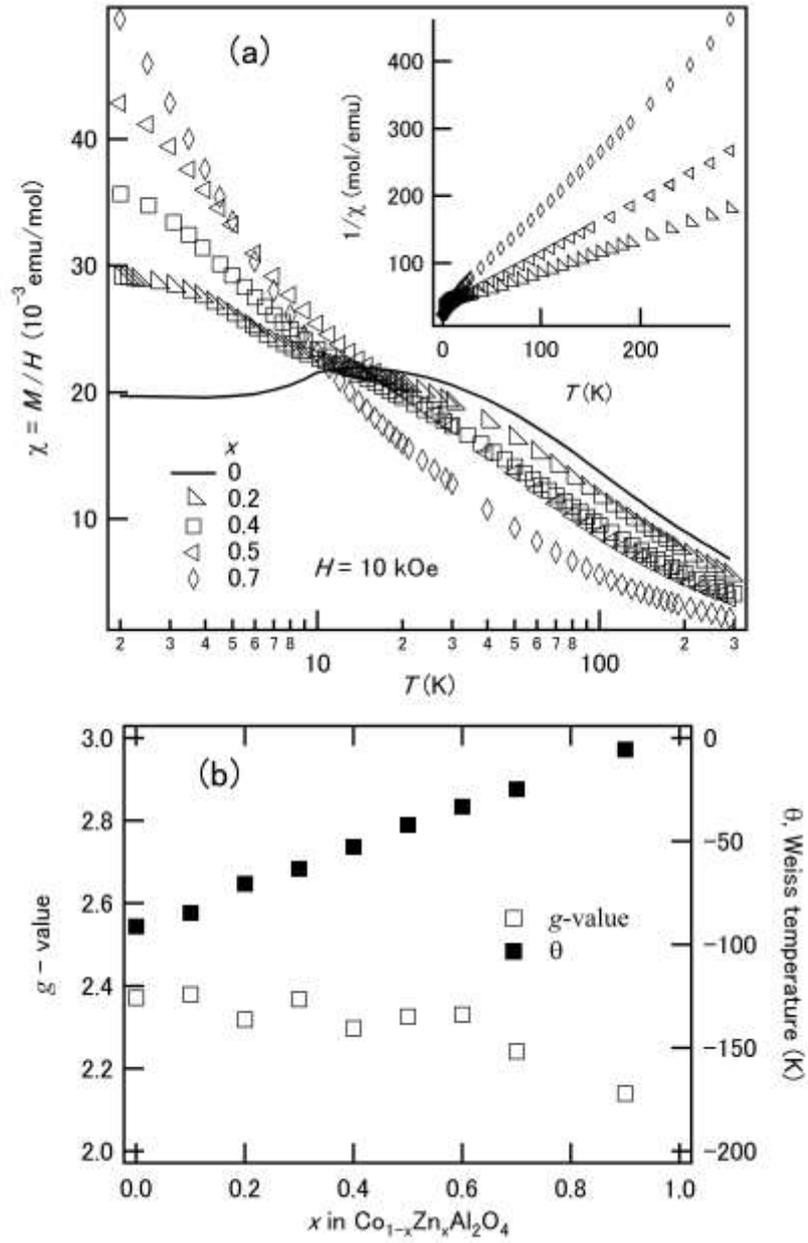

FIG. 5. (a) Magnetic susceptibility of $Co_{1-x}Zn_xAl_2O_4$ as a function of temperature measured at $H = 10$ kOe. The inset shows the temperature dependence of the reciprocal susceptibility. (b) The $x$-variations of the $g$-value and Weiss temperature.

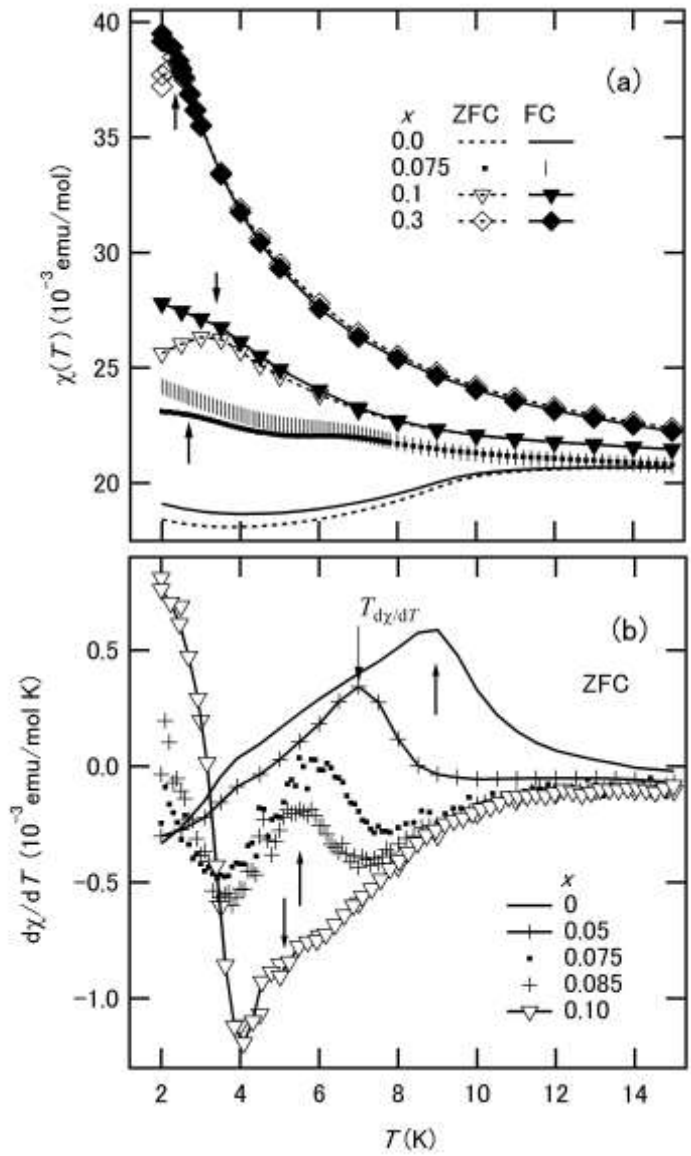

FIG. 6. (a) The temperature dependences of χ obtained at relatively low fields. Arrows indicate the cusp in χ(T) measured after the ZFC. (b) dχ/dT as a function of temperature below $x = 0.10$. The arrows indicate the anomalous points, $T_{d\chi/dT}$, at various Zn contents, $x$.

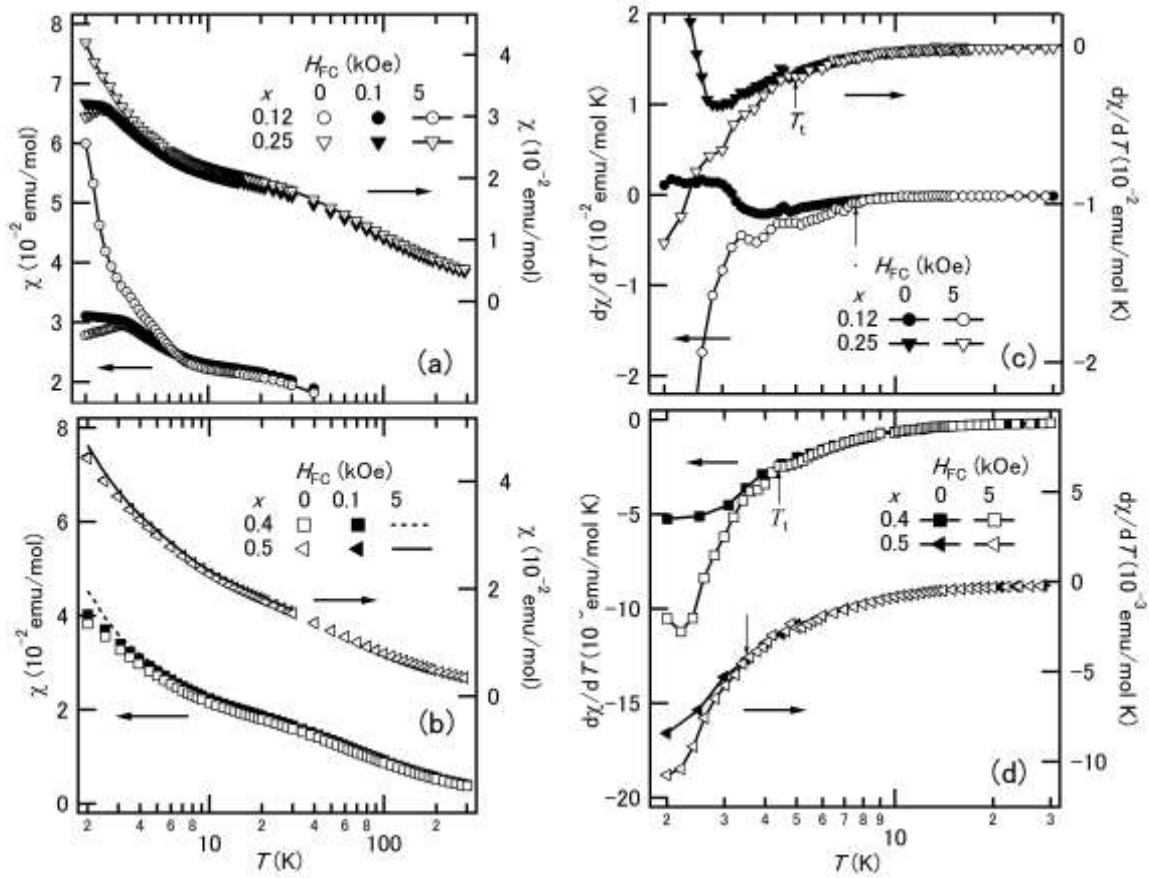

FIG. 7. (a) The temperature dependences of $\chi$ for (a) $x = 0.12$ and $0.25$ and (b) $x = 0.40$ and $0.50$ after the FC at $H_{FC} = 0$ (ZFC), $0.1$ and $5$ kOe. The $H_{FC}$ dependence of $d\chi/dT$ for (c) $x = 0.12$ and $0.25$ and (d) $x = 0.40$ and $0.50$. The vertical arrows indicate $T_t$ at various Zn contents, $x$.

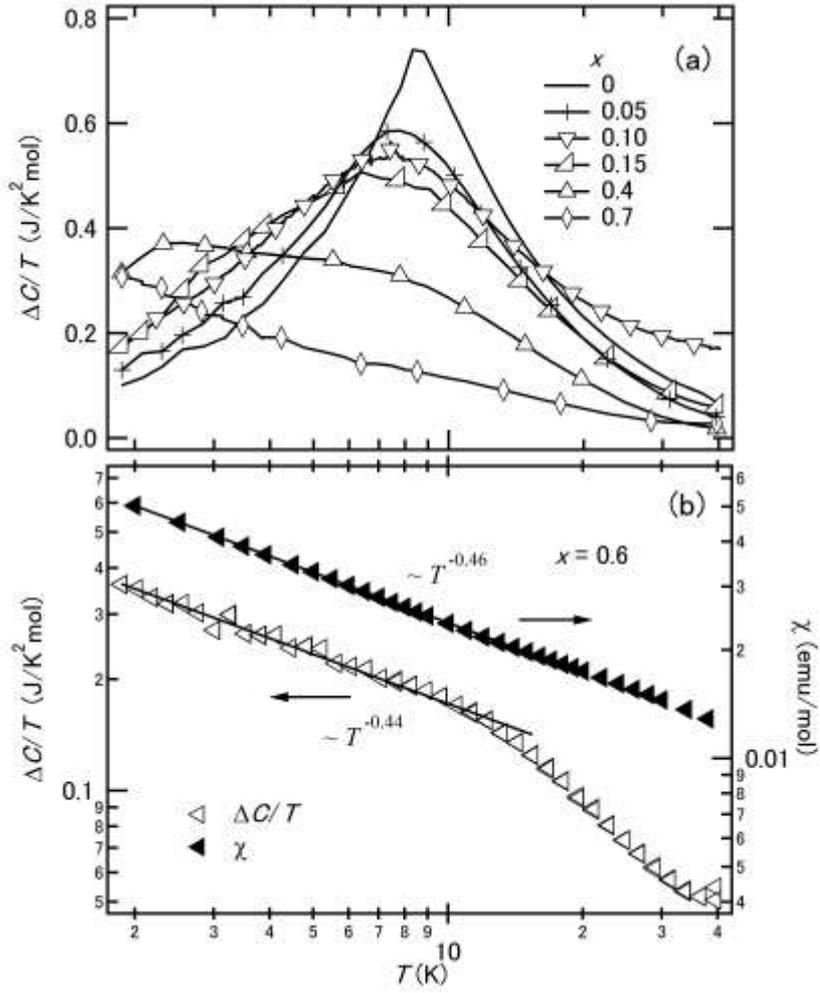

FIG. 8. (a) The temperature dependences of $\Delta C/T$ in $Co_{1-x}Zn_xAl_2O_4$ at zero field. (b) Open and filled triangles represent respectively $\Delta C/T$ and $\chi$ measured after ZFC as a function of temperature. Solid and dashed lines are the fitted curves below $T = 14$ K (see text for details).

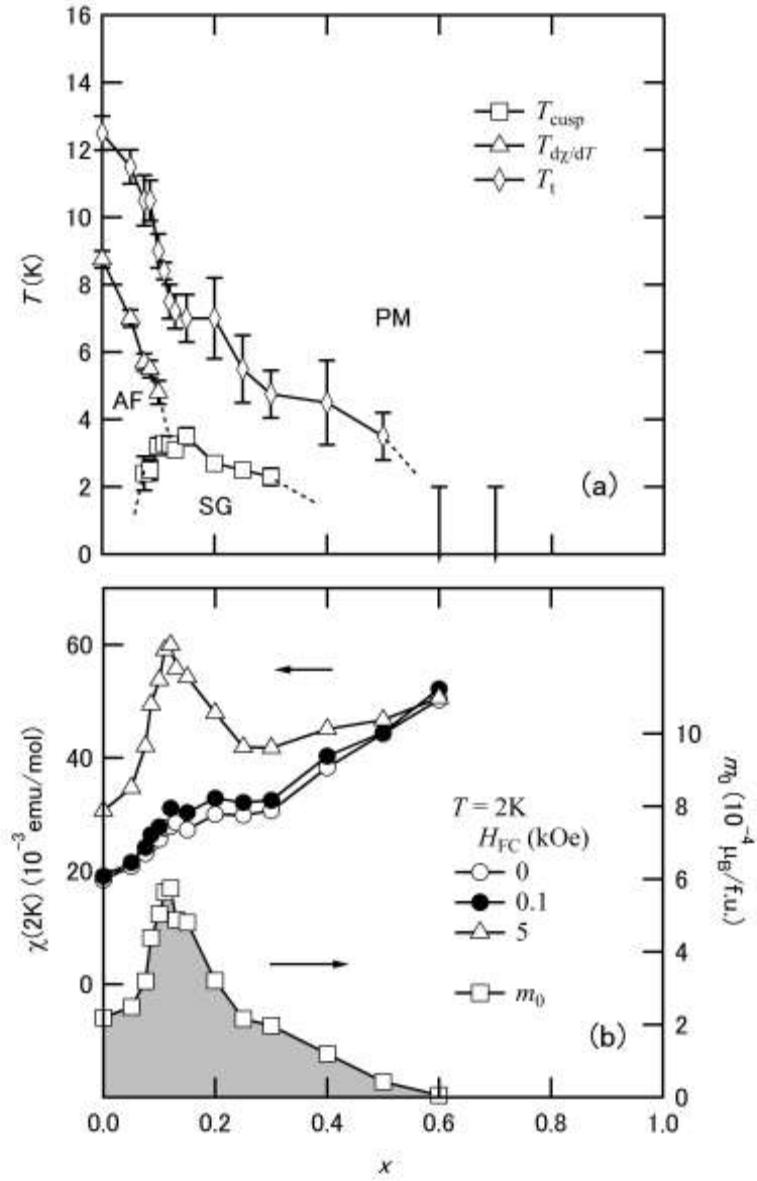

FIG. 9. (a) The transition temperatures, $T_{d\chi/dT}$, $T_{cusp}$ and $T_t$ are plotted on the $x$-$T$ plane. (b) The susceptibilities measured at $H = 0.1$ kOe after FC at $H_{FC} = 0$, 0.1 and 5 kOe and the field-induced magnetization, $m_0$(2 K, 5 kOe) are plotted as a function of $x$. Solid and dashed curves are aids to visualization.

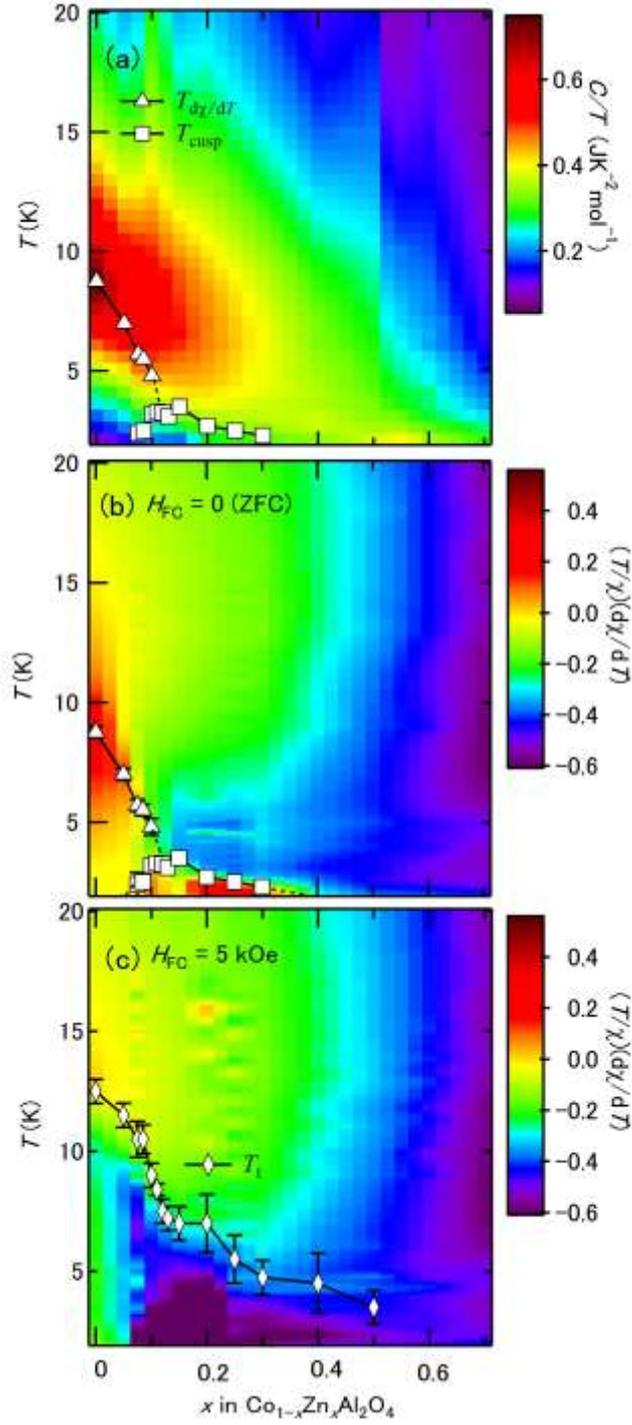

FIG. 10. False-color images of $C/T$ (a) and $(T/\chi)(d\chi/dT)$ measured after ZFC (b) and after FC under a magnetic field of $H_{FC} = 5$ kOe (c) on the $x$-$T$ plane, respectively. $T_{d\chi/dT}$, $T_{cusp}$ and $T_t$ are also plotted on the $x$-$T$ plane. Solid and dashed curves are aids to visualization.

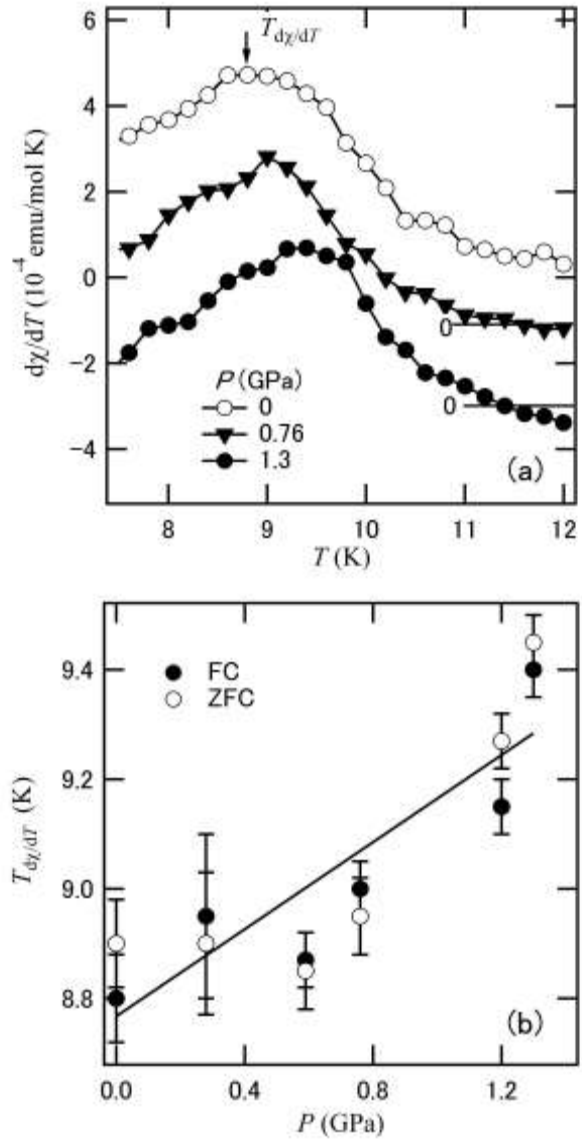

FIG. 11. Temperature dependence of $d\chi/dT$ at various pressures measured after ZFC (a) and $T_{d\chi/dT}$ as a function of pressure (b). $d\chi(T)/dT$ is offset at $P = 0.76$ and $1.3$ GPa. The solid line is a linear least-squares fit of $T_{d\chi/dT}$ with respect to pressure.

**Appendix**

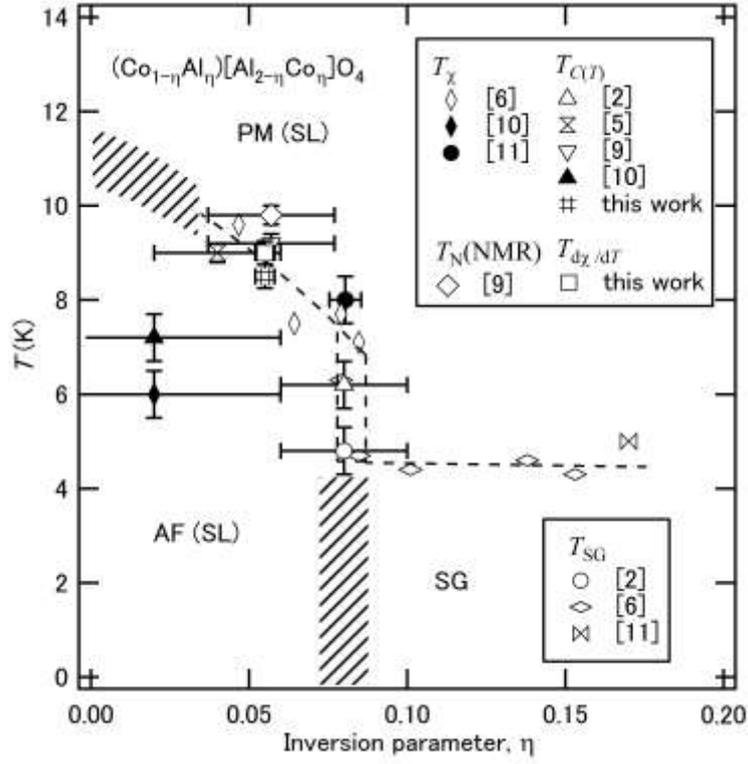

FIG. A1. Inversion parameter-temperature ($\eta$-$T$) phase diagram for $CoAl_2O_4$. Solid and open symbols represent the data points measured using single-crystal and polycrystalline samples, respectively. Dashed lines and hatched areas are aids for visualization based on the phase boundaries presented by Hanashima *et al.*[6] PM, AF, SL, and SG denote paramagnetic, antiferromagnetic, spin-liquid, and spin glass-like states, respectively. The numbers in brackets adjacent to the symbols are the references.